# An In-situ Annealing effect of Graphene-Graphene Interlayer Conduction

Jothiramalingam Kulothungan[1], Manoharan Muruganathan[*,1], and Hiroshi Mizuta[1,2]

[1] School of Materials Science, Japan Advanced Institute of Science and Technology, Nomi, Ishikawa 923-1292, Japan.

[2] Hitachi Cambridge Laboratory, Hitachi Europe Ltd., J.J Thomson Avenue, CB3 0HE Cambridge, United Kingdom

**ABSTRACT**

An interlayer distance modulation in twisted bilayer graphene is reported. This is achieved by an in-situ annealing technique. The transformation of systematic vacuum and hydrogen annealing effects in twisted bilayer CVD graphene on $SiO_2$ surface is reported based on experimental results. Incoherent interlayer conduction is observed in the twisted bilayer device. In-situ annealing efficiently removes the residues in the graphene-to-graphene interface and enhances the interlayer conduction. We demonstrate graphene-to-graphene interlayer resistance modulated by an order of magnetite at 5 K. We also report on the behavior of molecular hydrogen on graphene interlayer using the gate voltage-dependent resistance as a function of temperature at atmospheric pressure. It was observed that interlayer conduction in hydrogen/argon gas ambient is reduced. Results imply that modulation in the interlayer distance of graphene-to-graphene junction, as determined by the transport measurement investigation. Overall this result leads to the possibility of making electrically tunable devices using twisted bilayer graphene.

[*] Corresponding author. Tel: +81 761-51-1573. E-mail: mano@jaist.ac.jp (Manoharan Muruganathan)

# 1. Introduction

Bilayer graphene exhibits outstanding electronic properties based on how one layer of graphene is stacked on another graphene layer [1,2]. In AA stacked bilayer graphene, all carbon atoms of one layer of graphene are shifted in regards to the other layer by the interlayer distance, which results in two Dirac cones shifted in energy. In AB stacked bilayer graphene, half carbon atoms of one layer of graphene lie directly above the atoms of the second graphene layer and the remaining half carbon atoms lie over the centers of hexagons in the second layer, which results in a quadratic dispersion [3–6]. The new class of graphene bilayers, twisted bilayers, can be realised when the stacking order of the bilayer graphene is broken by an arbitrary rotational angle θ [7,8]. All the reported theoretical and the experimental works [9–23] now agree on the fact that twisted bilayer graphene shows distinct electronic properties based on the rotational angle between the two layers. The AA and AB bilayer graphene represent two extreme cases, corresponding to 0° and 60° rotational angle. The bilayer graphene crystal structure is symmetric for the 30° rotation [24]. The superposition of the twisted bilayer lattice periodicities produces the Moiré pattern. The Moiré superstructure is incommensurate for rotational angles greater than 10° with the graphite lattice leading to a strong suppression of the interlayer coupling. For the rotational angle of 30°, monolayer graphene electronic spectrum reappears [25]. Twisted bilayer graphene with a rotational angle less than 10° exhibits a strong enhancement in the interlayer coupling. More interestingly, the Moiré superstructure is commensurate at specific angles 30° ± 8.21°, 30° ± 2.20° and 30° ± 16.82° and produces the periodicity in the superlative structure [12,22].

Twisted bilayer graphene is of particular interest because of several exotic properties such as van Hove singularities [26], renormalization of Fermi velocity [27], and electronic localization [28,29]. Recently, many unique experimental characteristics were reported in twisted bilayer graphene including fractional quantum hall states [30], topological phase transition [31], Berry phase transition [32], and superconductivity [33–35]. These interesting properties are raised from the twist angle and the interlayer coupling in the twisted bilayer graphene. Modulating the interlayer coupling can change the physical characteristics of the system. Very recently, correlated phases were induced in twisted bilayer graphene without the so-called "magic angle" by varying the interlayer spacing by utilizing external hydrostatic pressure [36]. Thus, interlayer spacing and coupling play a crucial role in the twisted bilayer graphene. Unfortunately, many resist residues and other contaminations are inevitably left in the interlayer gap of twisted bilayer graphene during the fabrication process.

In this work, we present experimental evidence of improved graphene-to-graphene interlayer cleanness by in-situ annealing. We systematically investigate the transport properties of a twisted Chemical Vapor Deposition (CVD) bilayer graphene device by electrical measurements before and after in-situ annealing as well as in gas (Ar+$H_2$) ambient. We found that the interlayer resistance is reduced by an order of magnitude after in-situ annealing. This feature indicates that the interlayer distance of the graphene-to-graphene twisted bilayer graphene is modulated by in-situ annealing. When the measurement chamber pressure is increased to atmospheric pressure by Ar+$H_2$ gas the interlayer resistance increased. This is attributed to the interlayer distance modulation by the change in the pressure. In all these measurements, the interlayer resistance exhibits strong negative temperature dependence, which indicates the incoherent charge conduction between the bottom and the top graphene ribbons. These results indicate that, by modulating the interlayer distance, offering the strong evidence of the possibility of utilizing the more degrees of freedom for electronic applications using twisted bilayer graphene.

## 2. Experimental

### 2.1 Fabrication of CVD graphene twisted bilayer device

Fig. 1 illustrates the fabrication process for the graphene-graphene twisted bilayer device. First, the CVD grown graphene on Si/$SiO_2$ was used as the substrate (Fig. 1a). The contact metal electrode was fabricated by using the positive resist poly methylmethacrylate

(PMMA), and then the pattern was defined by electron beam (EB) exposure. The metal electrode (Cr/Au) was deposited by using the electron beam evaporation. The total thickness of the metal electrodes was 5/65 nm (Fig. 1b). The bottom graphene ribbon (BGNR) was fabricated using the high resolution negative resist hydrogen silsesquioxane (HSQ). The electron beam (EB) exposed HSQ turned into $SiO_2$ [37]. This HSQ resist mask was used to define the BGNR and the remaining graphene on the substrate was removed by oxygen plasma etching (Fig. 1c). After fabricating the BGNR, another layer of CVD grown graphene was transferred onto the sample using a wet chemical based transfer method (The graphene transfer method is given in the electronic supplementary information).

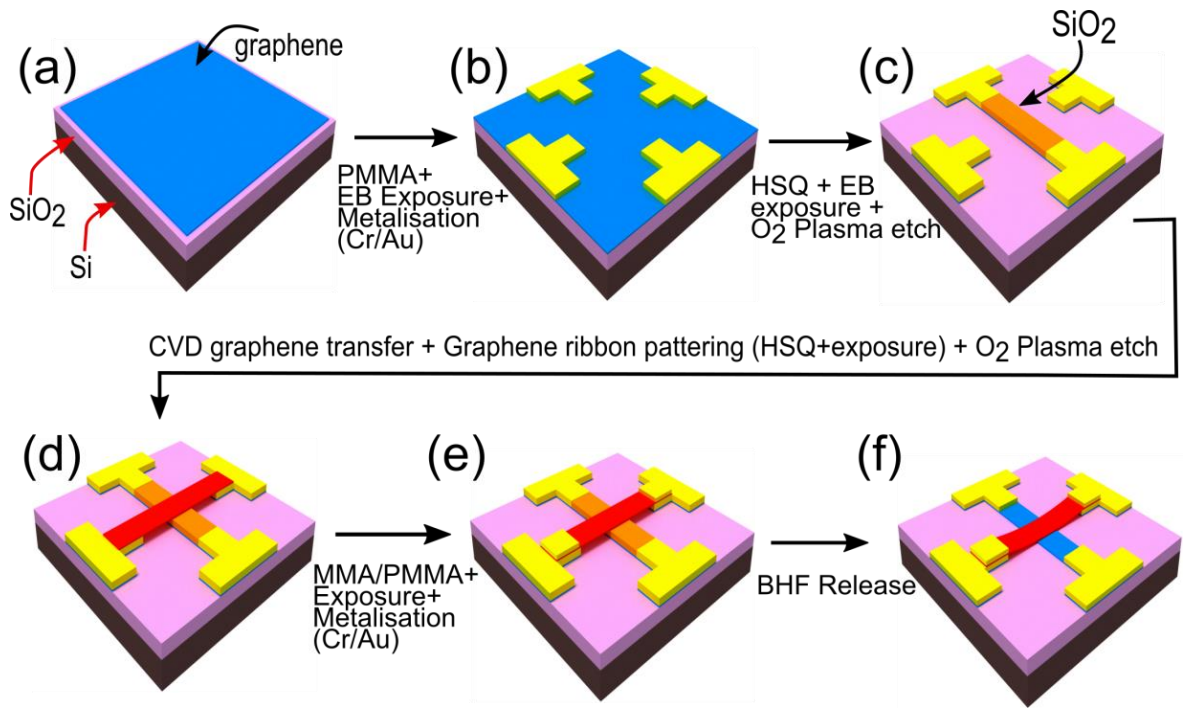

**Fig. 1.** The fabrication process of the twisted CVD bilayer graphene device. (a) CVD graphene substrate. (b) Contact electrode fabrication. (c) Bottom graphene ribbon with $SiO_2$. (d) Top graphene ribbon fabrication. (e) Anchor electrode fabrication for top graphene ribbon. (f) $SiO_2$ release using the BHF.

The top graphene ribbon (TGNR) was also fabricated using the same method described earlier for BGNR (Fig.1d). Unwanted graphene elsewhere on the sample was removed by using oxygen plasma. The TGNR was anchored tightly using metallization (Fig.1e). Finally, the sacrificial layer ($SiO_2$) in between the bottom graphene ribbon and the top graphene ribbon was etched in buffered hydrofluoric acid (1:5). This leads to realization of graphene-to-graphene interlayer (Fig. 1f). Thus, the removal of the $SiO_2$ between TGNR and BGNR leads to the graphene-to-graphene cross junction. The TGNR has a length of 1.8 µm and a width of 1 µm. The BGNR has a length of 3 µm and a width of 1 µm. The resultant graphene-to-graphene interlayer area is 1 µm × 1 µm. The schematic representation of CVD graphene twisted bilayer device shown in Fig. 2a.

*2.2 Device Electrical Characterization*

A semiconductor device analyzer (Keithley–4200 SCS) was used for all electrical measurements. Measurements were done in a variable temperature helium closed cycle cryocooler prober system, wherein it can reach the base temperature of T ≈ 5 K. Also, the measurement chamber can be vacuumed to ~ $10^{-4}$ Pa. To confirm the functionality of each graphene nanoribbon, the drain current was measured as a function of the back gate voltage. To investigate the interlayer transport characteristics of the device, the four-point probe method was used. Four contacts at the ends of the graphene ribbons were used as the electrical leads. For the interlayer transport measurement, we apply the current between BGNR and TGNR and the resultant voltage is measured at the remaining two terminals (Fig. 4a). The measurements were performed in a probe station equipped with in-situ annealing chamber connected by a load-lock valve (for more information see supplementary material). The measurements were carried out in the following order :- Step(1): measurement of the twisted CVD bilayer graphene device in the vacuum chamber (~$10^{-4}$ Pa) before annealing, Step(2): vacuum annealing was done at 200 °C with low pressure of ~$10^{-3}$ Pa for 2 hours followed by electrical measurements in the vacuum without exposing the device to the atmospheric environment, Step(3): hydrogen annealing was done at 300 °C in mixed gas ambient of Ar+$H_2$ (9:1) for 2 hours followed by electrical measurement in vacuum without exposing the device to an atmospheric environment, Step(4): electrical transport measurement in mixed gas ambient Ar+H$_2$ (9:1) at atmospheric pressure.

## 3. Results and discussion

*3.1 Raman spectroscopy of the CVD graphene twisted bilayer device*

The Raman spectra of the CVD graphene twisted bilayer device was measured using a Raman spectroscopy T64000 (HORIBA) with a solid-state crystal laser (λ = 532 nm) as the excitation source. To obtain the Raman spectrum precisely on the graphene-to-graphene overlapped area, laser beam spot was optimized to less than 1 µm by using optical lenses. Fig. 2b illustrates the scanning electron microscope image of the device. The twisted stacking of the graphene-to-graphene interlayer was demonstrated by Raman spectroscopy. Fig. 2c shows the CVD graphene twisted bilayer device spectrum relative to the monolayer CVD graphene spectrum on Si/SiO$_2$ substrate. The G peak appears at ~1585 cm$^{-1}$ and ~1589 cm$^{-1}$ respectively for twisted bilayer and monolayer. The bilayer graphene G peak has slightly larger FWHM (~2 cm$^{-1}$) while the peak location shows a slight red shift (~ 4 cm$^{-1}$) [38,39]. The 2D peak appears at ~2697 cm$^{-1}$ and ~2684 cm$^{-1}$ respectively for twisted bilayer and monolayer. 2D peak position of the twisted bilayer graphene is blue shifted relative to the 2D peak obtained for the monolayer graphene [40]. The relative shift of the 2D peak is ~13 cm$^{-1}$ (Fig. 2d). The 2D peak of the twisted bilayer graphene is shown in Fig. 2f. The 2D peak of the twisted bilayer graphene resembles a single Lorentzian peak similar to that of monolayer graphene (Fig. 2e), rather than multiple peaks as found in Bernal-stacked bilayer graphene. The Raman a spectrum of the twisted bilayer graphene strongly depends on twist angle between the graphene layers. By analyzing the difference in the Raman spectra from the twisted CVD graphene bilayer and the monolayer graphene regions in Fig. 2c an estimation of the twist angle θ is possible. This way the rotational angle of the twisted bilayer graphene was estimated to be ~ 10 degrees [40].

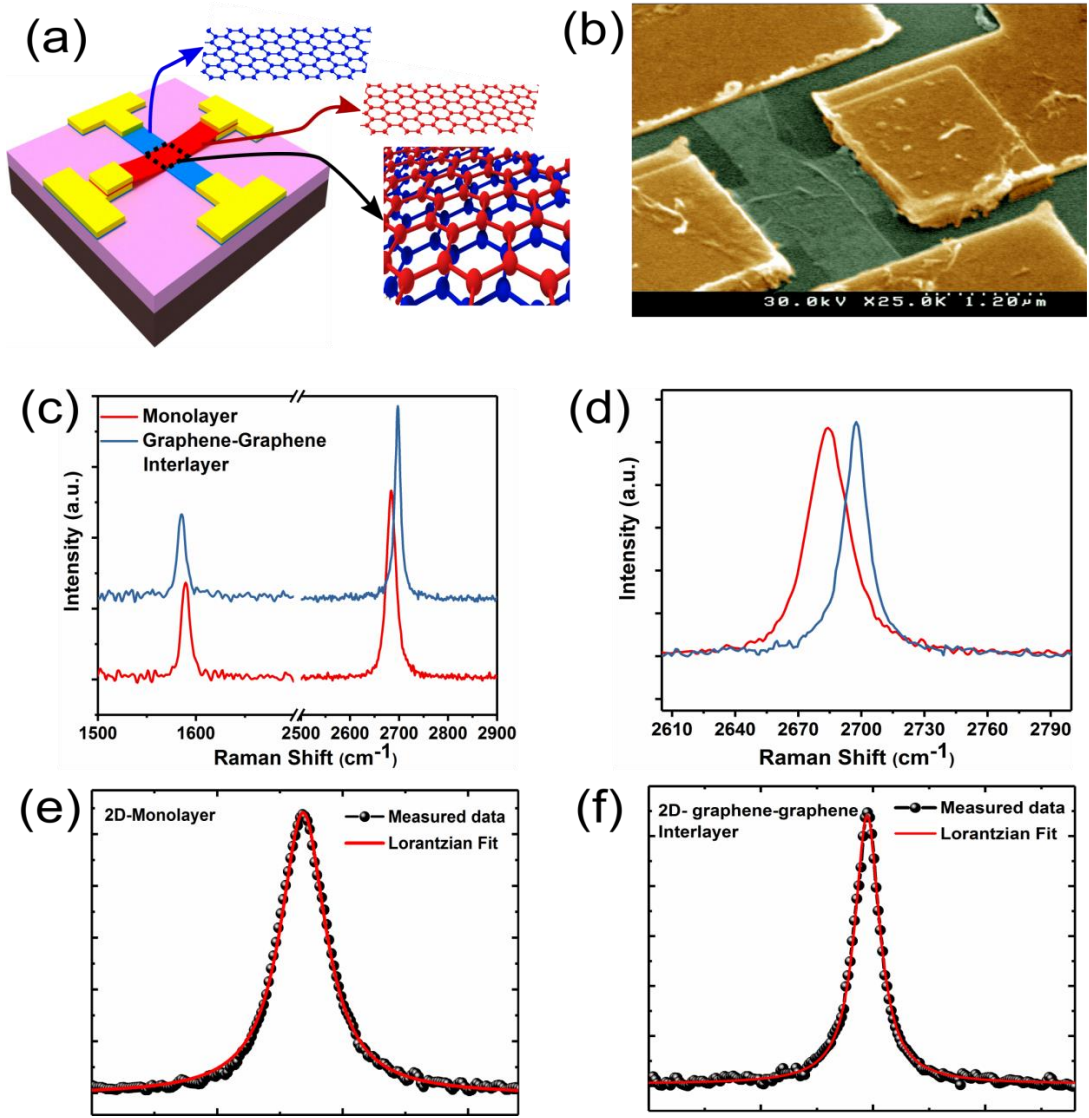

**Fig. 2.** (a) Schematic representation of CVD graphene twisted bilayer device. (b) SEM image of the fabricated and measured device (False colored). (c) Raman spectra were taken from the monolayer (CVD graphene substrate) and the twisted bilayer regions. (d) Double resonance (2D) peak of Raman spectra show the relative frequency shift of the 2D band in the twisted bilayer region with respect to the monolayer region. (e) The enlarged 2D band region of monolayer graphene in (c) with curves fitted by Lorentzian function. (f) The enlarged 2D band region of twisted graphene in (c) with curves fitted by Lorentzian function.

*3.2 Transport measurement results*

*3.2.1 Intra layer transport*

Four probe measurement configuration is shown in Fig. 3a. The current is applied between $V_S$ and $V_D$ terminals and the voltage drop across the interlayer is measured at $V_L$ and $V_H$ terminals respectively. The back gate characteristics of BGNR and TGNR before annealing (as fabricated) are shown in Fig. 3b. The back-gate characteristics of both the bottom and the top graphene ribbon are heavily p-doped. This p-doping mainly occurs due to the physisorbed molecules of $O_2$ and $H_2O$ onto the graphene surface or the water molecules present at the interface between GNR and $SiO_2$. When the back gate voltage Vg is applied, the electric fields for the BGNR and TGNR are different due to the screening of the gate induced electric field by BGNR. Fig. 3c shows the back gate characteristics of the BGNR and TGNR after the in-situ vacuum annealing. The charge neutrality point (CNP) is shifted to a negative gate voltage after in-situ vacuum annealing. Vacuum annealing effectively removes the water molecules and p-type dopants absorbed on the graphene surface and between the graphene layer and the $SiO_2$ substrate. Subsequently charged impurity scattering is also reduced, which is indicated by the sub-linear characteristics at high carrier density (Fig. 3b, TGNR). Moreover, these sub-linear characteristics depict the domination of short range and ripple scattering in the carrier transport rather than the charged impurities. Furthermore, the long duration of graphene device vacuum annealing gives n-type doping due to the oxygen deficiency induced surface state density in $SiO_2$ [41–43]. This is the origin of the negative shift in CNP after the vacuum annealing. After vacuum annealing, the device was hydrogen annealed for 3 hours at 275˚C. Hydrogen annealing more effectively cleans graphene when compared with vacuum annealing [44,45]. Fig. 3d shows the back gate characteristics of the BGNR and TGNR after the in-situ hydrogen annealing. The positive shift has been observed at the CNP for both graphene ribbons. It is known that hydrogen annealing of graphene devices leads to hydrogen intercalation at the graphene–$SiO_2$ interface. This process terminates the silicon dangling bonds of the $SiO_2$ surface on the substrate. Due to this hydrogen intercalation between BGNR and $SiO_2$ surface, CNP of BGNR is almost shifted closer to zero gate voltage. Remaining p-type dopants on TGNR and partially screening of electric field by the BGNR is attributed to the p-type doping observed from the TGNR. Compared to the as fabricated measurement results (Fig. 4b), we can observe the linear to sub-linear change in the characteristics around the gate voltage of ± 30 V for BGNR results. This demonstrates the removal of charged impurities and domination of short range and ripple scattering.

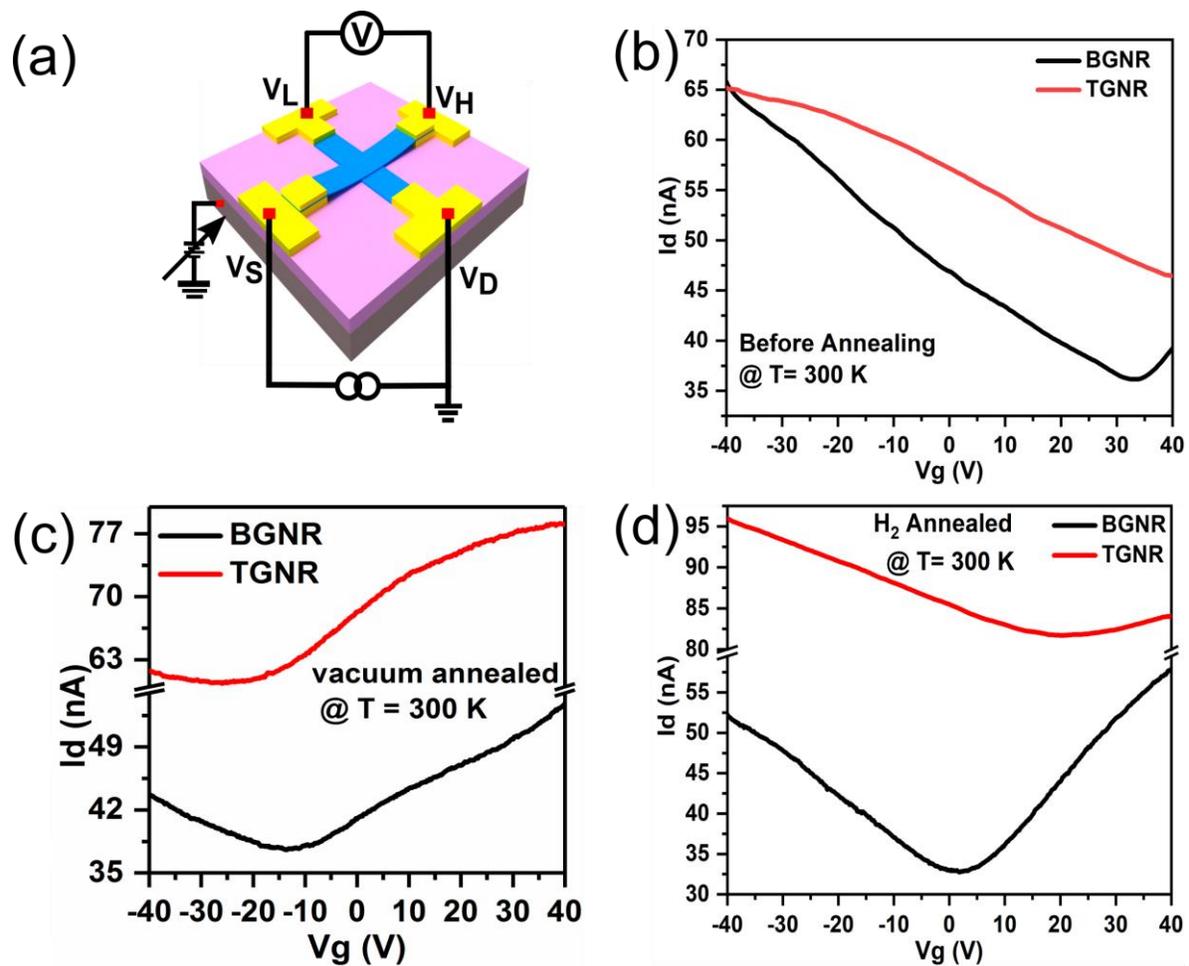

**Fig. 3.** (a) Measurement configuration of interlayer transport properties. (b) Back gate characteristics of the bottom ribbon and the top ribbon for the as fabricated sample. (c) After the vacuum annealing, (d) After the hydrogen annealing.

The temperature dependence backgate characteristics of BGNR and TGNR are shown in Fig. 4 for before annealing (as fabricated) and after hydrogen annealing conditions. As discussed before, the charge impurity scattering is reduced after the in-situ annealing process. Unlike GNR with a width less than 100 nm, conductance at CNP does not decrease above 30 % at low temperatures [46]. Also, the suppressed conductance region does not appear in these gate characteristics even at 5 K. These properties indicate that the bulk nature of individual GNRs are maintained in this twisted bilayer device. Despite these bulk characteristics, some oscillations appear at low temperatures. This can be attributed to the presence of charge puddles and grain boundaries in CVD graphene.

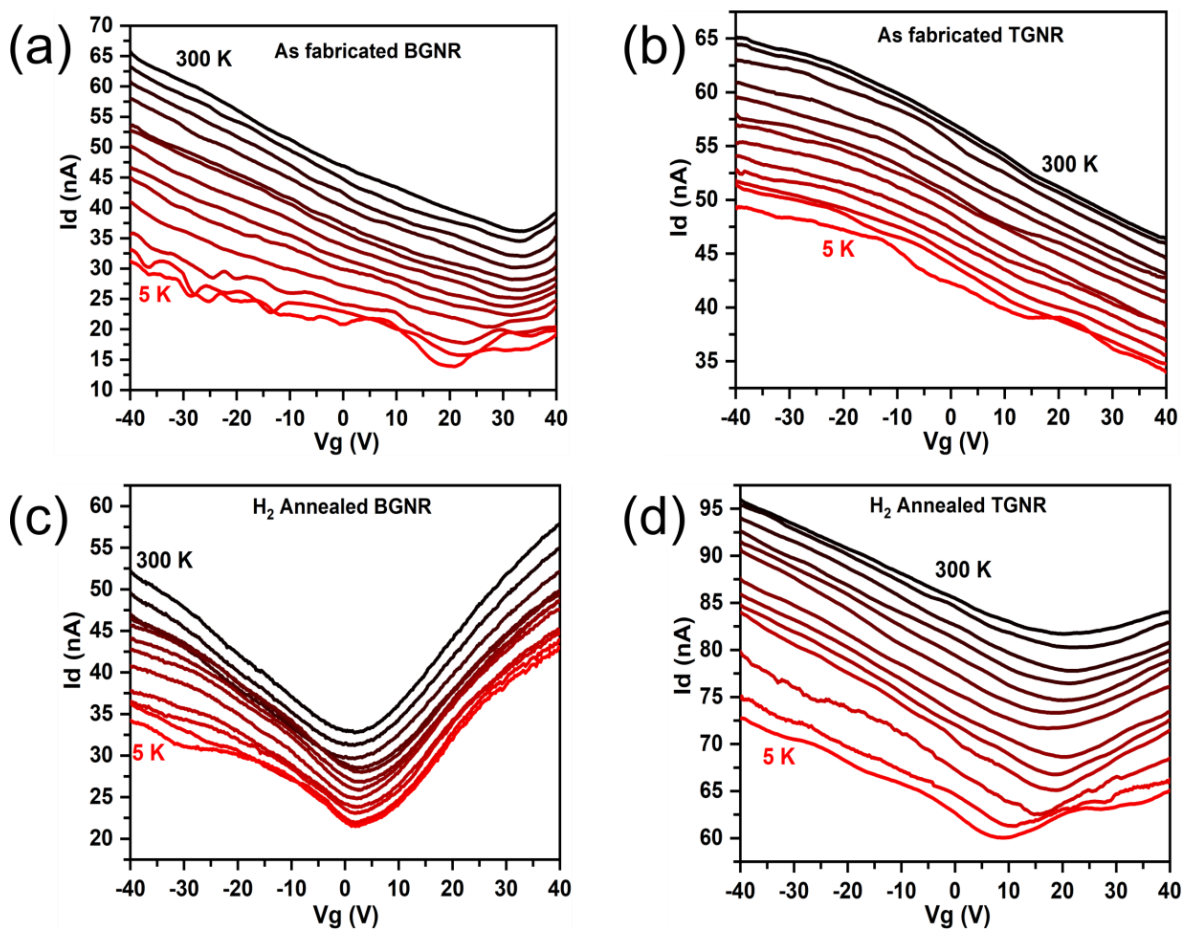

**Fig. 4.** Temperature dependence of back gate characteristics. (a-b) Temperature dependence backgate characteristics of the bottom graphene ribbon and the top graphene

ribbon for the as fabricated sample. (c-d) temperature dependence backgate characteristics of the bottom and the top graphene ribbon after in-situ hydrogen annealing. The color label is common for (a-d).

*3.2.2 Inter layer transport*

The interlayer resistivity measurement results of the twisted CVD bilayer graphene device are shown in Fig. 5. The gate voltage dependence of the interlayer resistivity at various temperatures is shown in Fig. 5a and 5b respectively for before in-situ annealing and after in-situ annealing. Individual BGNR and TGNR show linear I-V characteristics (Fig. 5c). The interlayer resistivity ($\rho_{inter}$) is defined as $R_{inter}*A/d$, where d is the interlayer distance (0.335 nm), A is the overlapping area between TGNR and BGNR, and $R_{inter}$ is the measured interlayer resistance using the four probe method. The as fabricated device shows 1100 $\Omega$ cm interlayer resistivity at 270 K. This is four orders higher than reported c-axis resistivity of ~ $10^{-1}$ $\Omega$ cm for highly oriented pyrolytic graphite (HOPG). The reported value of interlayer resistivity is ~ 4 X $10^{-3}$ $\Omega$ cm at 270 K for single crystalline graphite [47]. Intra-layer and interlayer resistivity decreases with the increase in the temperature as shown in Fig. 5d. These values of resistivity are extracted at CNP for both interlayer and intra-layer measurements. This negative intra-layer resistivity temperature dependence of CVD GNRs is slightly different from the moderate temperature dependence of exfoliated HOPG graphene [48]. This might be attributed to the presence of grain boundaries and other defects in CVD graphene. Moreover, this negative temperature dependence of intra-layer resistivity is distinct from the metallic behavior observed in single crystal graphite [47]. However, the semiconducting temperature dependence is observed for HOPG samples at around 270 K, and metallic behavior at low temperature [47]. Furthermore, intra-layer resistivity does not show the saturating trend at low temperature. However, interlayer resistivity decreases with increase in temperature from 100 K to 300 K and saturates to a constant value below 100 K. This trend is similar before and after annealing conditions (Fig. 5a-b). These results indicate the incoherent conduction in the twisted CVD bilayer graphene device similar to exfoliated twisted bilayer graphene [48].

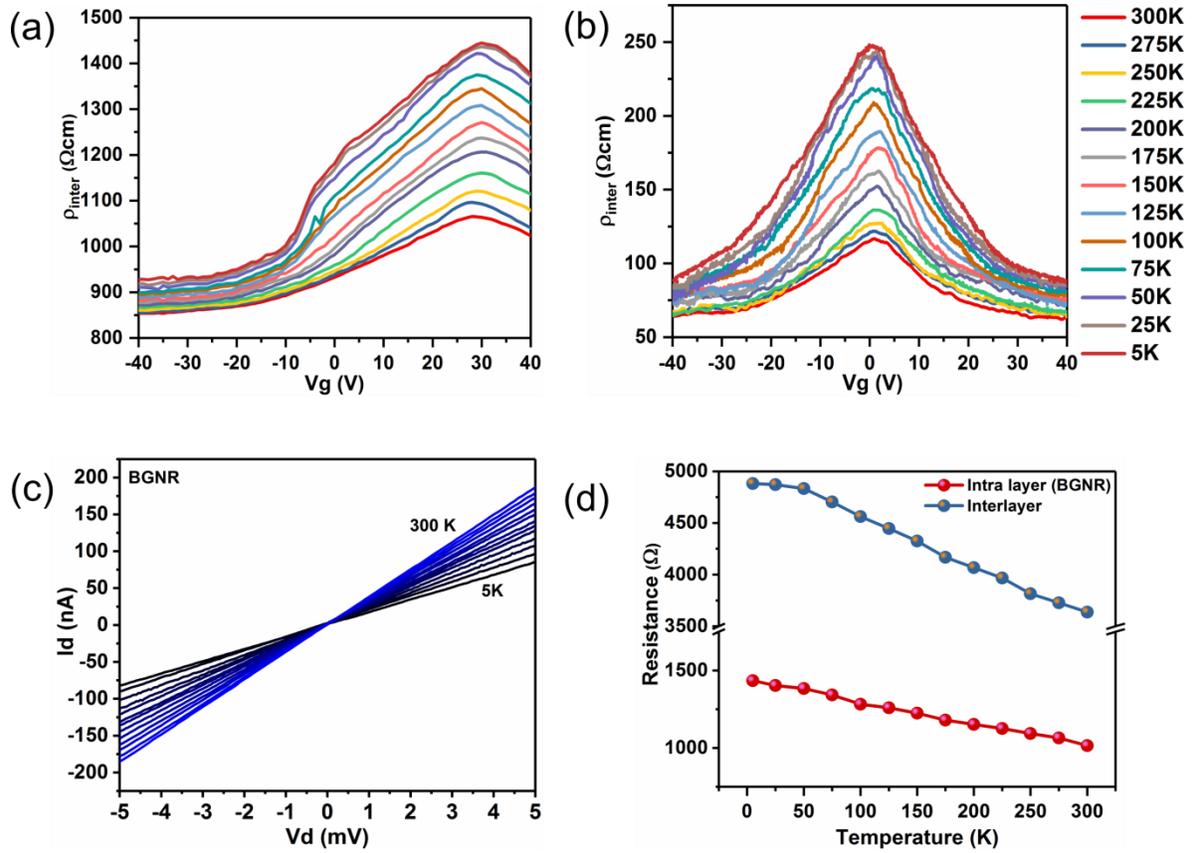

**Fig. 5.** (a) The interlayer resistivity of CVD graphene twisted bilayer device with a variation of the gate voltages (Vg) at different temperatures before annealing. (b) Interlayer resistivity of the device after in-situ annealing. (c) Current voltage characteristics of the intra-layer conduction at various temperatures from 5 to 300 K. (d) Temperature dependence of the intra-layer and interlayer resistance at CNP.

In-plane transport in graphene is bipolar in nature, which is observed in transfer characteristics of the graphene devices. An analogous observation in the interlayer transport between the graphene layers originates from two processes: (i) Metal-Insulator-Metal (MIM) model, density-of-states-dependent incoherent tunneling across the interlayer distance (van der Waals gap) with interlayer conductance

$G_{inter} = R_{inter}^{-1} \propto D_b(E_F)D_t(E_F)\Gamma$ [49,50] where $R_{inter}$, $D_{b,t}(E)$, $\Gamma$, and $E_F$ are the interlayer electrical resistance, energy-dependent density-of-states of the bottom and top graphene layers, interlayer transmission probability, Fermi energy of the graphene layers, respectively. Fig. 6a gives the schematic diagram of the MIM junction model with charge-puddle broadened DOS. In Fig. 6a, we show the normalized DOS for both electrons and holes in monolayer graphene. The solid lines represent the DOS in inhomogeneous system and the dashed solid lines represent the DOS in homogeneous systems, respectively. The electron band tail located at E < 0, induces the electron puddles at E < 0, whereas the hole band tail located at E > 0 gives rise to hole puddles at E > 0. Another process for interlayer conduction mechanism is (ii) interlayer conduction governed by electron-phonon scattering with $G_{inter} \propto nE_F^2/\omega_{ph}$ [51] where n, $E_F$, and $\omega_{ph}$ are the thermal population of the out-of plane beating phonon mode of interlayer, Fermi energy of the graphene layers, and phonon energy, respectively. Although both mechanisms lead to $G_{inter} \propto n$, in agreement with the observations for small *n*, they differ in their temperature dependences. As shown in Fig. 6c, the low temperature interlayer transport is temperature independent, which is consistent with incoherent quantum tunneling, whereas $R_{inter}$ decreases sharply for T >100 K. These characteristics suggest that the inception of phonon-assisted electrical conduction as the thermal population of interlayer phonons increases with increasing temperature. Both interlayer transport mechanisms are depended on thermal population of the phonon in a similar manner; the crossover temperature scale (~100 K) varies weakly with doping.

The effect of in-situ annealing is investigated in Fig. 6e which shows the temperature dependence of the interlayer resistivity before and after the in-situ annealing. Since the interlayer transport is dominated by the thermally-excited charge carriers over the potential barrier at high temperatures, the strong temperature dependence of the $\rho_{inter}$ is observed. At low temperatures, the interlayer carrier transmission occurs when the Fermi circles of the bottom and top layers intersect. As a consequence, interlayer conduction occurs through tunneling between two misoriented Fermi circles which leads to temperature-independent interlayer resistivity at low temperatures. The resistivity decreased to 250 Ω cm from 1445

Ω cm after in-situ annealing at 5 K. This behavior can be understood as follows, in-situ annealing at high temperature effectively removes the adsorbents at the interlayer. As a consequence, the interlayer distance between the twisted bilayer graphene is reduced. The interlayer distance is nothing but the tunneling width of a MIM model. Tunneling barrier width determines tunneling current density. The thinner the tunneling barrier width between the two graphene layers, the higher the tunneling current density, which is attributed to the larger transmittance coefficient. From this result, it is evident that interlayer distance is reduced by in-situ annealing.

*3.2.3 Ar+$H_2$ pressure-dependent electrical transport property*

The Ar+$H_2$ pressure dependent interlayer resistivity of twisted bilayer graphene device was measured from 300 K to 130 K at atmospheric pressure of Ar+$H_2$ gas in a measurement chamber (see supplementary material for details). The interlayer transport measurement in the gas ambient was started continuously after the in-situ annealing and the measurement was stopped at 130 K. The cryocooler system can't be operated below 130 K with the chamber in atmospheric pressure.

Fig. 6b shows the current-voltage characteristics of the intra-layer (BGNR) after $H_2$ exposure relative to I-V measured in vacuum (after in-situ annealing). We observed a reduction in in-plane resistance of the GNR upon $H_2$ exposure at 300 K. This can be attributed to electron doping due to $H_2$ dissociative adsorption [52,53]. This behavior can be explained as follows. During the reaction between molecular hydrogen and a graphene, hydrogen molecules dissociated on graphene can break C=C double bonds in the graphene structure. As a result of the process, two unpaired electrons are produced. An unpaired electron contributes to the formation of a C-H bond and the other is delocalized. The delocalized electrons induce a n-type doping effect on grapheme [54–56]. By further reducing the temperature, in-plane resistance gradually increased at low temperature as shown in the Fig. 6b inset.

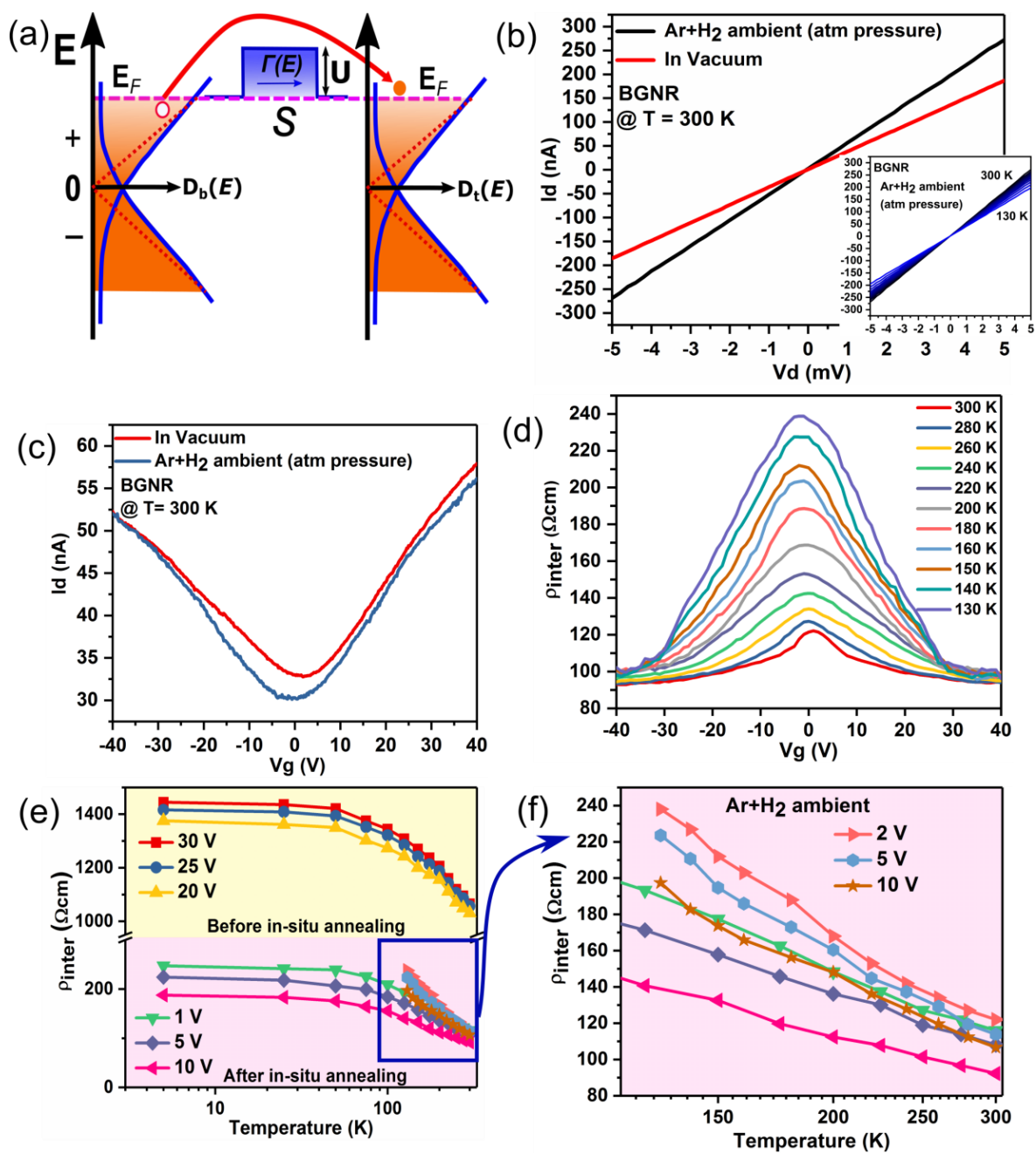

**Fig. 6.** (a) Schematic of the Metal-Insulator-Metal junction model with barrier height U and width S and individual layers are represented with charge-puddle broadened densities of states. (b) I-V characteristic of the intra-layer (BGNR) in Ar+H$_2$ ambient. Temperature dependence of I-V characteristics is shown in inset. (c) Back gate characteristic of the intra-layer (BGNR) in Ar+H$_2$ ambient. (d) The interlayer resistivity of CVD graphene twisted bilayer device with a variation of the gate voltages (Vg) at different temperatures in the Ar+H$_2$ gas ambient. (e) Temperature dependence of interlayer resistivity at various gate voltages. (f) Zoom-in view of (e) as indicated in the box.

The back gate characteristics of the intra-layer (BGNR) in Ar+H$_2$ are shown in the Fig. 6c in comparison with the transfer curve measured in vacuum at 300 K. n-type charge doping is induced in the H$_2$ ambient at atmospheric pressure, this is consistent with the results reported for the monolayer graphene [54]. The change in the CNP is ~ 1.5 V at 300 K. The gate voltage dependence of the interlayer resistivity in Ar+H$_2$ gas ambient at various temperatures is shown in the Fig. 6d.

The interlayer resistivity is increased to 125 Ω cm, ~ 6 % increase in H$_2$ ambient at 300 K. The interlayer resistivity is increases as the temperature decreases and it is not saturated at 130 K. We observed a shift in the resistivity maxima towards the negative gate voltage in a H$_2$ atmosphere and it is more prominent in the low temperature regime. Moreover, the interlayer resistivity increases linearly for measured temperature range as shown in Fig. 6f. The resistivity increased from 190 Ω cm to 240 Ω cm respectively for after in-situ annealing and in H$_2$ ambient at ~120 K. More specifically, the interlayer resistivity increased 26% at 130 K. The increase of interlayer resistivity indicates that the interaction between graphene layers was gradually reduced. The dissociative adsorption of H$_2$ molecules induces the decoupling of graphene interlayer, which results in an increase of the interlayer distance of twisted bilayer graphene. It is evident from our experimental results, that the interlayer distance of the twisted bilayer graphene is reduced significantly after the in-situ annealing process and increases after introducing the H$_2$ gas. Incoherent tunneling mechanism is observed in all the different cases.

## 4. Conclusion

We report the effects of vacuum and hydrogen annealing in the CVD graphene twisted bilayer graphene on SiO$_2$ surface by experimental results. Using the graphene-to-graphene twisted bilayer graphene device, we demonstrated the interlayer transport properties of the twisted bilayer graphene. It is found that in-situ annealing removes residues from the -fabrication and reduces the interlayer distance of the twisted graphene bilayer junction. Simultaneously, in-situ annealing enhances the interlayer conduction. In contrast, interlayer conduction is increased in the presence of Ar+$H_2$ gas ambient. Incoherent interlayer conduction is observed before and after the in-situ annealing. In-situ annealing method affects both intra-layer and interlayer transport. This systematic study clarifies the microscopic impact of vacuum and hydrogen annealing on CVD twisted bilayer devices, providing guidance for study the interlayer conduction mechanism in graphene based heterostructures.

**Acknowledgments**

This work was supported by the Grant-in-Aid for Scientific Research No. 25220904, 16K13650, and 16K18090 from Japan Society for the Promotion of Science and the Center of Innovation (COI) program of the Japan Science and Technology Agency. The authors thank Ahmed M.M. Hammam for his helpful discussions.